\documentclass{article}

\usepackage{amsmath, amsthm}
\usepackage{graphicx, xcolor}
\usepackage{dsfont}
\usepackage{tikz}
\usetikzlibrary{positioning}
\usepackage{amsfonts}
\usepackage{enumitem}   
\usepackage{xspace}
\usepackage{url}
\usepackage{soul}
\usepackage{caption}
\usepackage{subcaption}

\newcommand{\ind}[1]{\mathds{1}_{\{#1\}}}

\newcommand{\dd}{{\partial}}
\newcommand{\ee}{{\mathrm e}}

\newcommand{\figscale}{0.7}
\newcommand{\twopwidth}{0.49}

\graphicspath{{Figures/}}

\begin{document}

\newtheorem{mytheorem}{Theorem}
\newtheorem{myproposition}{Proposition}
\newtheorem{mylemma}{Lemma}
\newtheorem{myproof}{Proof}
\newtheorem{myalgorithm}{Algorithm}
\newtheorem{mycorollary}{Corollary}
\newtheorem{mydefinition}{Definition}
\theoremstyle{remark}
\newtheorem{myremark}{Remark}

\newcommand{\alggg}{hyper-scalable\xspace}
\newcommand{\algg}{scheme\xspace}
\newcommand{\alg}{hyper-scalable scheme\xspace}

\title{Optimal Hyper-Scalable Load Balancing with a Strict Queue Limit}

\author{
  Mark van der Boor\\
  Eindhoven University of Technology
  \and
  Sem Borst\\
  Eindhoven University of Technology
  \and
  Johan van Leeuwaarden\\
  Tilburg University, Eindhoven University of Technology
}

	\maketitle	
\date{}

\begin{abstract}
Load balancing plays a critical role in efficiently dispatching jobs in parallel-server systems such as cloud networks and data centers.
A fundamental challenge in the design of load balancing algorithms is to achieve an optimal trade-off between delay performance and implementation overhead (e.g.~communication or memory usage).
This trade-off has primarily been studied so far from the angle of the amount of overhead required to achieve asymptotically optimal performance, particularly vanishing delay in large-scale systems.
In contrast, in the present paper, we focus on an arbitrarily sparse communication budget, possibly well below the minimum requirement for vanishing delay, referred to as the hyper-scalable operating region. Furthermore, jobs may only be admitted when a specific limit on the queue position of the job can be guaranteed. 

The centerpiece of our analysis is a universal upper bound for the achievable throughput of any dispatcher-driven algorithm for a given communication budget and queue limit.
We also propose a specific hyper-scalable scheme which can operate at any given message rate and enforce any given queue limit, while allowing the server states to be captured via a closed product-form network, in which servers act as customers traversing various nodes.
The product-form distribution is leveraged to prove that the bound is tight and that the proposed hyper-scalable scheme is throughput-optimal in a many-server regime given the communication and queue limit constraints.
Extensive simulation experiments are conducted to illustrate the results.
\end{abstract}

\section{Introduction}
Load balancing provides a crucial mechanism for efficiently distributing jobs among servers in parallel-processing systems.
Traditionally, the primary objective in load balancing has been to optimize performance in terms of queue lengths or delays.
Due to the immense size of cloud networks and data centers \cite{duet14,MSY12,ananta13}, however, implementation overhead (e.g.~communication or memory usage involved in obtaining or storing state information) has emerged as a further key concern in the design of load balancing algorithms.
Indeed, the fundamental challenge in load balancing is to achieve scalability: providing favorable delay performance, while only requiring low implementation overhead in large-scale deployments.

The seminal paper~\cite{GTZ16} approached the above challenge by imposing the natural performance criterion that the probability of non-zero delay vanishes as the number of servers grows large.
It was shown that this can only be achieved with constant communication overhead per job when sufficient memory is available at the dispatcher.
There are in fact schemes which achieve a vanishing delay probability with only one message per job \cite{BB08,LXKGLG11,Stolyar15} or even fewer \cite{BBZ20}, but these rely on server-initiated updates as opposed to dispatcher-driven probes.
We defer a more extensive discussion of these papers and the broader literature
to a later stage in this introduction.

In the present paper we pursue the same intrinsic trade-off between performance and communication overhead, but focus on the optimal performance for a potentially scarce communication budget, and our perspective is fundamentally different in two respects.
First of all, we set the admissible message rate~$\delta$ to be arbitrary, and in particular to be far lower than one message per job, which we refer as the `hyper-scalable' operating regime.
This range is especially relevant in scenarios with relatively tiny jobs and a correspondingly massive arrival rate which may significantly exceed the message rate that can be sustained between the dispatcher and the servers, prohibiting even just one message per job.
Second, jobs may only be admitted when a strict limit~$K$ on the queue position of the job can be guaranteed.
This queue limit $K$ can have any value and is offered in systems of any size, as opposed to a zero queue length that is only ensured with high probability in a many-server regime.
The combination of a low communication budget per job and a strict admission condition is particularly pertinent for high-volume packet processing applications, where zero delay may not be feasible given the admissible message rate, but where an explicit queue limit is crucial.

As the cornerstone of our analysis, we establish a universal upper bound for the achievable throughput of any dispatcher-driven algorithm as function of~$\delta$ and~$K$, thus capturing the trade-off between performance and communication overhead.
We also introduce and analyze a specific hyper-scalable scheme which approaches the latter bound in a many-server regime, demonstrating that the bound is sharp.

\paragraph{Model set-up and hyper-scalable scheme}

We adopt the set-up of the celebrated supermarket model which has emerged as the canonical framework in the related literature (as further reviewed below), but add several salient features relevant for our purposes.
Specifically, we consider a system with $N$~identical servers of unit exponential rate and a single dispatcher where jobs arrive as a Poisson process of rate $N \lambda$.
The dispatcher is unaware of the service requirements of jobs and cannot buffer them, but must immediately forward them to one of the servers or block them. The throughput of the system is defined as the rate of admitted jobs per server. 

The blocking option is relevant since the dispatcher must enforce an explicit queue limit~$K$, and is only allowed to admit a job and assign it to a server if it can guarantee that the queue position encountered by that job is at most~$K$. 
Note that it is not enough for a job to end up in such a position thanks to a lucky guess, but that the dispatcher must have absolute certainty in advance that this is the case, and that a job must be discarded otherwise.
Discarding may be the preferred option in packet processing applications when handling a packet beyond a certain tolerance window serves no useful purpose.
In that case, processing an obsolete packet results in an unnecessary resource wastage and needlessly contributes to further congestion, and is thus worse than simply dropping the packet upfront. 

As mentioned above, the dispatcher is oblivious of the service requirements, which are exponentially distributed and thus have unbounded support. Hence, the dispatcher critically relies on information provided by the servers in order to enforce the queue limit~$K$, and is allowed to send probes for this purpose, requesting queue length reports at a rate $N \delta$.
In addition, the dispatcher is endowed with unlimited memory capacity, which it may use to determine which server to probe and when or to which server it will dispatch an arriving job.
Servers return instantaneous queue length reports in response to probes from the dispatcher, but are not able to initiate messages or send unsolicited updates when reaching a certain status.

With the above framework in place, we will construct a specific hyper-scalable scheme which is guaranteed to enforce the queue limit~$K$ and operate within the communication budget $\delta$.
The scheme toggles each individual server between two modes of operation, labeled open and closed.
An open period starts when the dispatcher requests a queue length update from the server and the reported queue length is below~$K$;
during that period the server is not working, and waits for incoming jobs from the dispatcher, seeing its queue only grow.
Once the queue length reaches the limit~$K$, a closed period starts, ending when the dispatcher requests the next update after $\tau$ time units;
during that period the server is continuously working as long as jobs are available, without receiving any further jobs, thus draining its queue.
When the queue length reported at an update is exactly~$K$, the open period has length zero, and the next closed period starts immediately.
By construction, the above-described mechanism maintains a queue limit of~$K$ at all times and induces a message rate of at most~$1/\tau$ per server, which makes $\tau=1/\delta$ the obvious choice.

\paragraph{Main contributions}
The main contributions of the paper may be summarized as follows.
First of all, we establish a universal upper bound $\lambda^*(\delta, K)$ for the achievable throughput of any dispatcher-driven algorithm subject to the communication budget per server in terms of~$\delta$ and the queue limit~$K$.
The upper bound relies on a simple yet powerful argument which counts the number of jobs that can be admitted per message given the queue limit~$K$ and the message rate~$\delta$.
While the macroscopic view of the argument covers a broad range of strategies with possibly dynamic and highly complex update rules, the nature of the upper bound strongly points to the superior properties of constant update intervals.

Armed with that insight, we propose a hyper-scalable scheme which can operate at any given message rate~$\delta$ and enforce any given queue limit~$K$.
At the same time, the scheme is specifically designed to produce system dynamics that can be represented in terms of a closed product-form queueing network, in which the servers act as customers traversing various nodes. This furnishes tractable expressions for the relevant stationary distributions and in particular the blocking probability.
The expression for the blocking probability is used to prove that the achieved throughput approaches the minimum of the above-mentioned upper bound and the normalized job arrival rate $\lambda$ in a many-server regime.
This in turn demonstrates that the upper bound is tight and that the proposed hyper-scalable scheme provides optimality in the three-way trade-off among queue limit, communication and throughput.

\paragraph{Background on load balancing algorithms}
Load balancing algorithms can be broadly categorized as static (open-loop), dynamic (closed-loop), or some intermediate blend, depending on the amount of state information (e.g. queue lengths or load measurements) that is used in dispatching jobs among servers.
Within the category of dynamic policies, one can further distinguish between
dispatcher-driven (push-based) and server-oriented (pull-based) approaches.
In the former case, the dispatcher `pushes' jobs to the servers and takes the initiative to collect state information for that purpose, while the servers play a passive role and only provide state information when explicitly requested.
In contrast, in server-oriented approaches, the servers may pro-actively share state information with the dispatcher, and indirectly `pull' in jobs by advertising their availability or load status.
The use of state information naturally allows dynamic policies to achieve better performance, but also involves higher implementation complexity
(e.g.~communication overhead and memory usage) as mentioned earlier.
The latter issue has emerged as a pivotal concern due to the deployment of large-scale cloud networks and data centers with immense numbers of servers handling massive amounts of service requests.

The celebrated Join-the-Shortest-Queue (JSQ) policy provides the gold standard in the category of dispatcher-driven algorithms and offers strong stochastic optimality properties.
Specifically, in case of identical servers, exponentially distributed service requirements and a service discipline at each server that is oblivious to the actual service requirements, the JSQ policy achieves minimum mean delay among all non-anticipating policies~\cite{EVW80,Winston77}.
In order to implement the JSQ policy, however, a dispatcher relies on instantaneous knowledge of the queue lengths at all the servers, which may involve a prohibitive communication burden, and may not be scalable. Related is the join-below-threshold scheme \cite{ZWTSS17}, which is throughput-optimal, but the dispatcher-driven variant is not scalable either.

The latter issue has spurred a strong interest in so-called JSQ($d$) strategies, where the dispatcher assigns incoming jobs to a server with the shortest queue among $d$~servers selected uniformly at random.
This involves $d$ message exchanges per job (assuming $d \geq 2$), and thus drastically reduces the communication overhead compared to the full JSQ policy when the number of servers~$N$ is large. 
At the same time, even a value as small as $d = 2$ yields significant performance improvements in the many-server regime $N \to \infty$ compared to purely random assignment ($d = 1$) \cite{Mitzenmacher01,VDK96}.
This is commonly referred to as the “power-of-two” effect.
Similar power-of-$d$ effects have been demonstrated for heterogeneous servers, non-exponential service requirements and loss systems in \cite{BLP10,BLP12,MKM15,MKMG15,MM14a,XDLS15}.

Unfortunately, JSQ($d$) strategies lack the ability of the conventional JSQ policy to achieve zero queueing delay as $N \to \infty$ for any finite value of~$d$.
In contrast, if $d$ grew with~$N$, making it possible to drive queueing delay to zero~\cite{MBLW16-3, LY18}, the communication overhead would grow unboundedly.
A noteworthy exception arises for batches of jobs when the value of~$d$ and the batch size grow suitably large, as can be deduced from results in~\cite{YSK15}.
Leaving batch arrivals aside though, it is in fact necessary for $d$ to grow with~$N$ in order to achieve zero queueing delay, since results in the seminal paper~\cite{GTZ16} show that this is fundamentally impossible with a finite communication overhead per job, unless memory is available at the dispatcher to store state information.

The latter feature is exactly at the core of the so-called Join-the-Idle-Queue (JIQ) scheme \cite{BB08,LXKGLG11}, where servers advertise their availability by transferring a ‘token’ to the dispatcher whenever they become idle, thus generating at most one message per job.
The dispatcher assigns incoming jobs to an idle server as long as tokens are outstanding, or to a uniformly at random selected server otherwise.
Remarkably, the JIQ scheme has the ability of the full JSQ policy to drive the queueing delay to zero as $N \to \infty$, even for generally distributed service requirements~\cite{FS17,Stolyar15}.

Note that for no single value of~$d$, a JSQ($d$) strategy can rival the JIQ scheme which simultaneously provides low communication overhead and asymptotically optimal performance.
As alluded to above, this superiority reflects the power of server-oriented approaches in conjunction with memory at the dispatcher.
The value of memory in load balancing was already studied
in~\cite{AGL10,Mitzenmacher02} in a `balls-and-bins' context.
Related work in~\cite{Mitzenmacher00} examines how much load balancing performance degrades when delayed information is used.
A framework for mean-field analysis for JSQ($d$) strategies with memory is developed in~\cite{LN13}.
The authors of~\cite{AD20} use mean-field limits to determine the minimum required value of~$d$ for JSQ($d$) strategies with memory to achieve zero queueing delay. The possibilities with limited memories are explored in \cite{HV20}.

As described above, the main interest in the line of work sparked by \cite{GTZ16} has focused on the amount of communication overhead and/or memory usage required to drive the queueing delay to zero in a many-server regime.
While there are known schemes to achieve that with just one message per job,
even that may still be prohibitive, especially when jobs do not involve big computational tasks, but tiny data packets which each require little processing.
In such situations the sheer message exchange in providing queue length information may be disproportionate to the actual amount of processing required.
While the overhead can be reduced, that only appears feasible for sufficiently low load~\cite{BBZ20}, and it remains largely unknown what the best achievable performance is for a given communication budget below one message per job.
Motivated by these issues, we focus on dispatcher-driven schemes that can operate at an arbitrarily low communication budget, and that can additionally enforce a specific queue limit for every admitted job.
To the best of our knowledge, this hyper-scalable perspective has not been pursued before, with the exception of~\cite{BBL19} which however does not consider explicit queue limits or optimality properties.

The literature on load balancing algorithms has ballooned in recent years, and the above discussion provides a non-exhaustive cross-section with some of the classical paradigms and results most pertinent to the present paper.
We refer to~\cite{BBLM18} for a more comprehensive survey discussing related job assignment mechanisms, further model extensions and alternative asymptotic regimes (e.g. heavy-traffic and non-degenerate slow down scalings).

\paragraph{Organization of the paper}
The main results are presented in Section \ref{sec:max}: the upper bound for the throughput and the analysis of the \alg, using a closed queueing network. In Section \ref{sec:num} we provide simulation results to further illustrate the behavior of the \alg. An extension of the \alg that also aims to minimize queue lengths is introduced in Section \ref{sec:ext}. In Section \ref{sec:proof} we establish product-form distributions for a general closed queueing network scenario which covers both the hyper-scalable scheme and the latter extension as special cases. We conclude with some remarks and suggestions for further research in Section \ref{sec:concl}.

\section{Main results}\label{sec:max}
In this section we discuss the main results, which can be summarized as follows. There is a function $\lambda^*$ of $\delta$ and $K$, such that subject to a message rate $\delta$ and queue limit $K$,
\begin{itemize}
\item the throughput of any dispatcher-driven algorithm is bounded from above by $\min\{\lambda^*(\delta,K), \lambda\}$,
\item the throughput of our \alg approaches $\min\{\lambda^*(\delta,K), \lambda\}$ as $N\to \infty$.
\end{itemize}
These two results are covered in Subsections \ref{subsec:bound} and \ref{subsec:alg}, respectively.

\subsection{Universal upper bound}\label{subsec:bound}
We establish the upper bound for a slightly more general scenario with heterogeneous server speeds. Denote the speed of the $n$-th server by $\mu_n$ for $n=1,\hdots,N$.
The next theorem shows that the achievable throughput of any dispatcher-driven algorithm subject to the message rate~$\delta$ and queue limit~$K$ is bounded from above by
\[
\lambda^*(\delta, K) = \delta M_K(\bar\mu / \delta),
\]
with
\begin{equation}\label{eq:MKT}
M_K(\tau) = \sum_{k = 0}^{K - 1} (1 - \alpha_k(\tau)),
\end{equation}
$\alpha_k(\tau) = \ee^{- \tau} \sum_{i = 0}^{k} \frac{\tau^i}{i!}$ and $\bar\mu = \frac{1}{N} \sum_{n = 1}^{N} \mu_n$ denoting the system-wide average server speed.
Note that $M_K(\tau)$ may be equivalently written as
\[
M_K(\tau) = K - \sum_{k = 0}^{K - 1} (K - k) \ee^{- \tau} \frac{\tau^k}{k!},
\]
and may be interpreted as the expected value of the minimum of~$K$ and a Poisson distributed random variable with mean $\tau$. \\

\begin{mytheorem}\label{th:gen}
The expected number of jobs that any dispatcher-driven algorithm can admit subject to the queue limit~$K$ during a period of length $T_0$ with at most $\delta N T_0$ message exchanges cannot exceed $2 K N + \lambda^*(\delta, K) \times N T_0$, for any $\delta>0$.
In particular, the achievable throughput with a message rate of at most $\delta>0$ is bounded from above by $\lambda^*(\delta, K)$.
\end{mytheorem}

Recall that we defined throughput as the rate of admitted jobs per server, and note that the throughput is naturally bounded from above by the normalized arrival rate $\lambda$.

\begin{proof}
As noted earlier, since the execution times are exponentially distributed and thus have unbounded support, the dispatcher relies on information provided by the servers in order to enforce the queue limit~$K$.
Specifically, the dispatcher earns `passes' for admitting $k$~jobs when a server reports $k=0,\hdots,K$~service completions since the previous update, and cannot admit any job without relinquishing a pass.
Thus, the number of jobs that the dispatcher can admit during a particular time period cannot exceed the sum of (i) the maximum possible number of $K N$ passes initially available; (ii) the maximum possible number of $K N$ passes earned at the first update from each server during that period, if any; and (iii) the number of additional passes obtained at further updates over intervals that fell entirely during that period, if any.
Now suppose that the dispatcher requests $L_n$ queue length reports from the $n$-th server during a period of length~$T_0$, one after each of the update intervals of lengths $T_{n,1}, \dots, T_{n,L_n}$, with $\sum_{l = 1}^{L_n} T_{n,l} \leq T_0$ for all $n = 1, \dots, N$ and $L = \sum_{n = 1}^{N} L_n \leq \delta N T_0$.
Then the number of passes earned at the $l$-th update equals the number of service completions during the time interval~$T_{n,l}$, which depends on the queue length at the start of that interval. However, this random variable is stochastically bounded from above by when the queue was full with $K$ jobs at the start of the interval. In the latter case the number of passes earned is given by the minimum of~$K$ and a Poisson distributed random variable with parameter $\mu_n T_{n,l}$. 
We deduce that the expected total number of passes obtained at all these updates is bounded from above by
\begin{equation}
\sum_{n = 1}^{N} \sum_{l = 1}^{L_n} M_K(\mu_n T_{n,l}),
\label{sum1}
\end{equation}
and to prove the first statement of the theorem it thus remains to be shown that this quantity is no larger than $\lambda^*(\delta, K) \times N T_0$.
It is easily verified that
\[
\frac{\dd^2 M_K(t)}{\dd t^2} = - \ee^{-t} \frac{t^{K - 1}}{(K - 1)!} < 0,
\]
implying that $M_K(t)$ is concave as function of~$t$.
As an aside, the above expression may be intuitively explained from the fact that the first derivative $\frac{\dd M_K(t)}{\dd t}$ equals the probability that exactly $K - 1$ unit-rate Poisson events occur during a period of length~$t$, while the (negative) derivative of the latter probability equals that very same probability by virtue of the Kolmogorov equations for a pure birth process.
Because of concavity, we obtain that \eqref{sum1} is no larger than $L \times M_K(\tau)$, with
\begin{equation}
\tau = \frac{1}{L} \sum_{n = 1}^{N} \mu_n \sum_{l = 1}^{L_n} T_{n,l} \leq \frac{1}{L} \sum_{n = 1}^{N} \mu_n T_0 = \bar\mu \frac{N T_0}{L}.
\label{sum2}
\end{equation}
Invoking the fact that $\frac{\dd M_K(t)}{\dd t} > 0$, i.e., $M_K(t)$ is increasing in~$t$, we may write
\begin{equation}\label{eq:derlam}
L \times M_K(\tau) \leq L \times M_K\left(\bar\mu \frac{N T_0}{L}\right) = \lambda^*(\gamma, K) \times N T_0,
\end{equation}
with $\gamma = \frac{L}{N T_0} \leq \delta$.
It is easily verified that
\begin{equation}\label{eq:incr}
\frac{\partial \lambda^*(x, K)}{\partial x} = K - K \ee^{- 1 / x} \sum_{k = 0}^{K} \frac{(1/x)^k}{k!} > 0,
\end{equation}
i.e., $\lambda^*(x, K)$ is increasing in~$x$, and hence $\lambda^*(\gamma, K) \leq \lambda^*(\delta, K)$, which completes the proof of the first statement of the theorem.

Finally, to prove the second statement, we consider the long-term scenario $T_0\to\infty$. The number of jobs that are admitted per time-unit per server then equals $(2 K N + \lambda^*(\delta, K) \times N T_0)/(N T_0) \to \lambda^*(\delta, K)$ and the message rate per server equals at most $\delta N T_0 / (N T_0) = \delta$.
\end{proof}

\paragraph{Properties of $\lambda^*$}\label{subsec:prop}
We now state some properties of $\lambda^*(\delta, K)$ and discuss their consequences, where we assume without loss of generality that $\bar\mu=1$. In the next subsection we will introduce a \alg which is able to achieve this throughput in the many-server regime. For now, we will reflect the properties in light of the maximum throughput that is possible for any dispatcher-driven load balancing algorithm given a message rate $\delta$.

\begin{myproposition}\label{prop:props}
 $\lambda^*(\delta,K)$ has the following properties:
\begin{enumerate}[label=\emph{(\roman*)}]
\item $\lambda^*(\delta,K)$ is strictly increasing in both $\delta$ and $K$,
\item $\lambda^*(\delta,K) \uparrow 1$ as $\delta \to \infty$, 
\item $\lambda^*(\delta,K) \downarrow 0$ and $\lambda^*(\delta, K) / \delta \to K$ as $\delta \downarrow 0$,
\item for $a\leq 1$, $\lambda^*(a/K, K) \to a$ as $K \to \infty$. 
\end{enumerate}
\end{myproposition}

\begin{proof}
$\lambda^*(\delta,K)$ is strictly increasing in $\delta$ because of \eqref{eq:incr}
and is strictly increasing in $K$ since $1-\alpha_k(\tau)>0$ in \eqref{eq:MKT}. 
For Properties (ii) to (iv), note that 
\[
\begin{split}
&\delta[K-K \ee^{-1/\delta}-\ee^{-1/\delta}((K-1) /\delta+(K-2)(1/\delta)^{y(\delta)}) ]\\ &\leq
\delta(K-\ee^{-1/\delta}\sum_{i=0}^{K-1} (K-i) \frac{(1/\delta)^i}{i!} )=
 \lambda^*(\delta,K) \leq \min(\delta K,1),
 \end{split}
\]
with $y(\delta)=2$ when $\delta\geq 1$ and $y(\delta)=K-1$ when $\delta< 1$. All limiting statements are true for the LHS and RHS of the previous equation, therefore proving these properties for $\lambda^*(\delta,K)$ too.
\end{proof}

\begin{figure}[h]\centering
\includegraphics[width=\figscale\columnwidth]{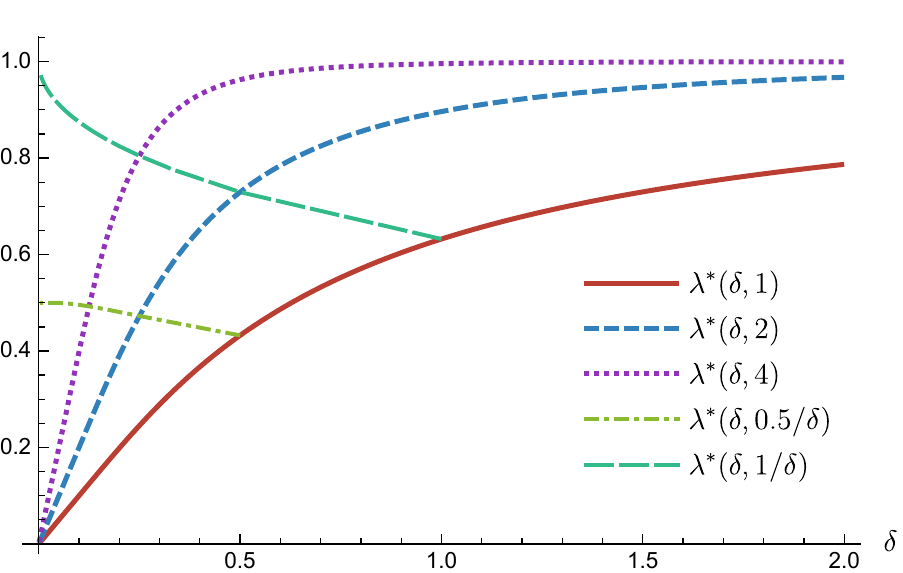}
\caption{Visualization of the throughput bound $\lambda^*(\delta,K)$ for various values of $K$ as function of $\delta$. For the fourth and fifth graph, only values of $\delta$ are evaluated for which the second argument is integer-valued.} \label{fig:lambdastar}
\end{figure}

The properties in Proposition \ref{prop:props} are visualized in Figure \ref{fig:lambdastar}. They can be interpreted intuitively and practically too. For Property (i), when the communication budget is expanded, i.e. $\delta$ is increased, more jobs can be dispatched to queues that are guaranteed to be short. Similarly, more jobs can be admitted into the system if the queue limit is raised, i.e., $K$ is increased. Property (i), in conjunction with Theorem \ref{th:gen}, implies that a throughput $\lambda^*(\delta,K)$ cannot be achieved with a message rate strictly below $\delta$, or a queue limit strictly below $K$.

Property (ii) shows that as the message rate grows large, full server utilization can be achieved. With an unlimited message rate, the dispatcher is able to find idle servers immediately, a necessary requirement for achieving full server utilization irrespective of the queue limit $K$.

Property (iii) shows that, first, when no communication is allowed, no jobs can be sent to queues that are guaranteed to be short. The further specification of the limit indicates that $K$ jobs are admitted into the system per message. This in turn reveals that when the communication is extremely infrequent, all messages result into finding an idle server, and thus provide the dispatcher with $K$ passes to admit jobs.

Finally, Property (iv) is somewhat similar to Property (iii). When the queue limit $K$ increases, one needs fewer messages in order to achieve a server utilization level $a$. With $a=1$, Property (iv) shows that one message per $K$ jobs is needed in order to achieve full server utilization, which is a somewhat similar conclusion as the one from Property (iii).

\subsection{The \alg}\label{sec:alg}\label{subsec:alg}
We now introduce the \alg in full detail for the case of homogeneous servers.

At all times, the dispatcher remembers the most recent queue length that was reported by every server. Furthermore, the dispatcher records the number of jobs that have been sent to every server since the last update from that server. When the sum of these two numbers is strictly less than the queue limit $K$, a server is labeled open, and otherwise closed.

Whenever a job arrives to the dispatcher, it is assigned to an open server, if possible. There are two options for how to select an open server. Either an open server is selected uniformly at random (random case), or the open server that was interacted with (i.e. updated or received job) the longest ago is selected (FCFS case). The job is dropped when no open servers exist.

Exactly $\tau$ time units after a server was labeled closed, the dispatcher will request a queue length update of the server. The server becomes open when this queue length is strictly less than $K$, and the server remains closed for another $\tau$ time units if the queue length equals $K$, in which case the dispatcher will request the next queue length update after another $\tau$ time units.
The \alg is a dispatcher-driven algorithm, since only the dispatcher initiates messages and every server can track itself when it is labeled open by the dispatcher: exactly when the sum of the queue length during the latest update and the number of jobs received since then is strictly below $K$. \\

Note that by construction the \alg respects the queue limit $K$ at all times and involves a message rate of at most $1/\tau$. In addition, the scheme has been specifically designed to allow explicit analysis and derivation of provable capacity benchmarks. As it turns out, a crucial feature in that regard is for the servers to refrain from executing jobs while being marked open. This feature ensures that the queue length is exactly $K$ at the moment a server becomes closed. The average number of job completions in an interval of length $\tau$ then equals $M_K(\tau)$, so one message leads to $M_K(\tau)$ admitted jobs on average, immediately yielding the following result.
\begin{mycorollary}\label{cor:comm}
The average number of messages per admitted job is equal to
$1 / M_K(\tau)$,
regardless of $\lambda$ and $N$. 
\end{mycorollary}

While the forced idling of servers during open periods may seem inefficient, the next theorem shows that the proposed hyper-scalable scheme is in fact throughput-optimal in large-scale systems, given the message rate $\delta$ and queue limit $K$, with the choice $\tau = 1/\delta$.

\begin{mytheorem}\label{th:asymopt}
For any $\delta>0$, the throughput achieved by the \alg with $\tau = 1/\delta$ approaches $\min\{\lambda^*(\delta,K), \lambda\}$ as $N\to \infty$. 
\end{mytheorem}
Since the \alg obeys the queue limit $K$ and involves a message rate of at most $\delta$, Theorems \ref{th:gen} and \ref{th:asymopt} combined imply that it is throughput-optimal as $N\to \infty$.

According to Theorem \ref{th:gen} and Property (i) of Proposition \ref{prop:props}, one would require a message rate of at least $\delta$ to achieve a throughput of $\lambda^*(\delta,K)$. Theorem \ref{th:asymopt} shows that the throughput of the \alg approaches $\lambda^*(\delta,K)$ as $N\to \infty$ when $\lambda \geq \lambda^*(\delta,K)$. A combination of these two observations (and the fact that $\lambda^*(\delta,K)$ is continuous in $\delta$) indicates that the message rate of the \alg must approach $\delta$ as $N \to \infty$ when $\lambda \geq \lambda^*(\delta, K)$.
This in turn implies that the expected duration of an open period must become negligible, compare to the length $\tau$ of a closed period, i.e. the fraction of time that a server is marked open vanishes. \\

We now proceed with an outline of the proof of Theorem 2. 

\paragraph{Analysis}
For brevity, a server is said to be in state $k$ when the sum of the queue length at its latest update epoch and the number of jobs the server has received since, equals $k$. This means that all servers in state $k < K$ are labeled open and servers in state $K$ are labeled closed.
In view of the homogeneity of the servers, it is useful to further introduce
$N(t) = (N_0(t), N_1(t), \hdots, N_{K-1}(t), N_K(t))$, with $\sum_k N_k(t) = N$, where $N_k(t)$ stands for the number of servers in state $k$ at time $t$. While the vector $N(t)$ provides a convenient representation, it is worth emphasizing that it does not provide a Markovian state description.

We now explain how individual servers transition between various states. When a job arrives to the system, the state of an open server will change from $k<K$ to $k+1$. An update of a server may cause the server to change state too. The new state of the server equals the number of jobs that are left in queue after the update interval of $\tau$ time units. The number of jobs that were served follows a truncated Poisson distribution, so the probability $p_k$ that exactly $k$ jobs remain, equals $p_k := \ee^{-\tau}\frac{\tau^{K-k}}{(K-k)!}$ for $k>0$ and $p_0 := 1-\ee^{-\tau}\sum_{i=0}^{K-1} \frac{\tau^i}{i!}$. 
When $k<K$ jobs are left, the state of the server becomes $k$. When there are $K$ jobs left, the state of the server does not change and remains $K$.

It is important to observe that service completions of jobs do not cause direct transitions in server states. The reason is twofold. When a server is open, it stops working on jobs, so there are no such completions at open servers. For closed servers, all servers are aggregated; the number of jobs in queue is not taken into account. Only after the period of length $\tau$, the number of jobs left in queue is determined indirectly by using the transition probabilities as specified above.\\

Although the vector $N(t)$ does not provide a Markovian state description as noted above, its evolution can be described in terms of a closed queueing network, in which the servers act as customers in the network, traversing various nodes corresponding to their states. Specifically, the closed queueing network consists of one multi-class ``single-server'' node with service rate $\lambda N$ in which the customers can be of classes $0,1,\hdots, K-1$, and one ``infinite-server'' node with deterministic service time $\tau$ that holds all class-$K$ customers. A service completion at the single-server node makes one customer transition. The class of the customer changes from $k$ to $k+1$ if $k<K-1$, or the customer transitions to the infinite-server node if its class was $K-1$. When multiple customers are present at the single-server node, the customer that transitions is either selected uniformly at random (random case), or the customer that has been in the single-server node for the longest time is selected (FCFS case). 
Finally, upon a service completion at the infinite-server node a customer moves to the single-server node as class $k<K$ with probability $p_k$, or directly re-enters the infinite-server node with probability $p_K$.

\newcommand{\pp}{\gamma}
\newcommand{\qq}{\kappa}
\newcommand{\PP}{p}
\newcommand{\customer}{customer\xspace}
\newcommand{\customers}{customers\xspace}
\newcommand{\Customer}{Customer\xspace}
\newcommand{\Customers}{Customers\xspace}

\begin{figure}[h!]
\centering
\includegraphics[width=0.6\columnwidth]{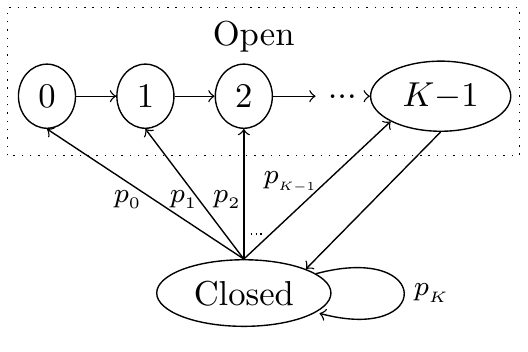}
\caption{Schematic representation of the circulation of an individual customer in the closed queueing network.}\label{fig:repr}
\end{figure}

A schematic representation is shown in Figure \ref{fig:repr}. We define $\pp_k$ as the relative throughput value of class-$k$ customers. With $\pp_K=1$, it follows that $\pp_k = p_0+\hdots+p_k = 1-\alpha_{K-1-k}(\tau)$ for $k<K$.

By virtue of the above-described equivalence, the process $N(t)$ representing the server states under the \alg inherits the product-form equilibrium distribution of the closed network as stated in the next proposition.
\begin{myproposition}\label{prop:productform}
The equilibrium distribution of the system with $N$ servers is 
\begin{equation}\label{eq:productform}
\pi(n_0,n_1,\hdots, n_{K-1},n_K) =
G_N^{-1}
\frac{(n_0+\hdots+n_{K-1})!}{n_0! \dots n_{K-1}!}
\left(\prod_{i=0}^{K-1} \left(\frac{\pp_i}{\lambda N}\right)^{n_i}\right)
\frac{\tau^{n_K}}{n_K!}
\end{equation}
if $n_0 +\hdots+n_K = N$, with normalization constant
\[
G_N=
\sum\limits_{v_0+\hdots+v_{K-1}+w = N} \frac{(v_0+\hdots+v_{K-1})!}{v_0! \dots v_{K-1}!} \left(\prod_{i=0}^{K-1} \left(\frac{\pp_i}{\lambda N}\right)^{v_i}\right)
\frac{\tau^{w}}{w!}.
\]
\end{myproposition}

A proof of Proposition \ref{prop:productform} can be found in Section \ref{sec:proof}, and the product-form equilibrium distribution may be informally understood as follows. The infinite-server node allows a product-form distribution even for deterministic service times. While traditionally exponentially distributed service times are considered, the equilibrium distribution is insensitive to the service time distribution at the infinite-server node and only depends on its mean, see Section \ref{sec:proof} for details.
As mentioned above, the service discipline at the single-server node with exponentially distributed service times may either be FCFS or random order of service.
In the case of the FCFS discipline, albeit not being reversible \cite{kelly_book_rev}, the single-server node with multiple classes can be represented as an order-independent queue \cite{BKK95,K11}. According to Theorem 2.2 in \cite{K11}, the queue is quasi-reversible, which is sufficient for a product-form distribution.
For random order of service, which is a symmetric service discipline, the single-server node is reversible, yielding a product-form as well.

The equilibrium distribution \eqref{eq:productform} can be simplified when only the number of open and closed servers matters. This immediately yields an expression for the blocking probability $L_N$ as provided in the next corollary. 

\begin{mycorollary}\label{cor:baseline}
The equilibrium probability of there being $n$ open servers and $N-n$ closed servers under the \alg equals
\begin{equation}\label{eq:eqprobopenclosed}
\pi(n, N-n) = \sum_{n_0+\hdots+n_{K-1}=n} \pi(n_0,\hdots,n_{K-1},N-n) =  
\frac{
\left(\frac{M_K(\tau)}{\lambda N}\right)^n \frac{\tau^{N-n}}{(N-n)!}
}{
\sum_{w=0}^N\left(\frac{M_K(\tau)}{\lambda N}\right)^w \frac{\tau^{N-w}}{(N-w)!}
}.
\end{equation}
In particular, because of the PASTA property, the blocking probability is given by 
\begin{equation}\label{eq:blockprob}
L_N = \pi(0,N) = \frac{
\frac{ \left(x N \right)^N }{N!}
}
{
\sum_{w=0}^N \frac{\left(x N\right)^w}{w!}
},
\end{equation}
with $x = \lambda \tau/ M_K(\tau) = \lambda / \lambda^*(1/\tau,K)$.
Finally,
\[
L_N \overset{N\to\infty}{\to} \max\{0, 1-\lambda^*(1/\tau,K) / \lambda\}.
\]
Specifically, $L_N\downarrow 0$ as $N\to \infty$ when $\lambda \leq \lambda^*(1/\tau, K)$.
\end{mycorollary}

Suppose that the allowed message rate is $\delta$ as stated in Theorem \ref{th:asymopt}, then put $\tau=1/\delta$. When $\lambda\leq \lambda^*(1/\tau,K)$, the blocking probability vanishes in the many-server regime according to Corollary \ref{cor:baseline}, and thus the throughput approaches $\lambda$. When $\lambda > \lambda^*(1/\tau, K)$, the acceptance probability tends to $\lambda^*(1/\tau,K)/\lambda$ and the throughput approaches $\lambda \times \lambda^*(1/\tau,K) / \lambda = \lambda^*(1/\tau,K)$. These two statements combined yield Theorem \ref{th:asymopt}.\\

Theorem \ref{th:asymopt} allows us to equivalently view $\lambda^*(\delta,K)$ as the throughput that is achieved by the \alg as $N\to\infty$ when $\lambda \geq \lambda^*(\delta,K)$. We now revisit properties (ii) and (iii) as stated in Proposition \ref{prop:props} from that perspective. In the limiting scenario $\delta \to \infty$, $\tau \downarrow 0$, servers are updated after an infinitesimally small time, which in turn alerts the dispatcher immediately when even a single job has been served. This ensures that all servers can work at full capacity.

In the scenario $\delta \downarrow 0$, $\tau \to \infty$, update periods become extremely long. Every update that does happen, will definitely find an idle server and allow for $K$ admitted jobs, explaining why $\lambda^*(\delta,K) \approx K \delta$ for small $\delta$. 

\begin{myremark}
Note that with the queue limit~$K$ in force we may assume each server to have a finite buffer of size~$K$.  In case of a finite buffer, the queue limit~$K$ would automatically be enforced, even if the dispatcher were allowed to forward jobs without any advance guarantee.  With the option of "(semi)-blind guesses", the throughput bound would trivially become~$1$ (the average server speed), and Property (iii) indicates that the achievable throughput $\lambda^*(\delta, K)$ without lucky guesses could be (substantially) lower when $\delta$ is (significantly) smaller than $1/K$.  However the throughput of~$1$ can only be approached for a high arrival rate, at the expense of severe blocking, whereas the hyper-scalable scheme can deliver the throughput $\lambda^*(\delta, K)$ with negligible blocking asymptotically.
\end{myremark}

\section{Simulation experiments and optimality benchmarks}\label{sec:num}
In this section we conduct various simulation experiments to further benchmark the properties of the \alg and make several comparisons. Throughout we set the queue limit $K=2$, yielding the throughput bound $\lambda^*(\delta,2)=2\delta-2\delta \ee^{-1/\delta}-\ee^{-1/\delta}$ as function of the message rate $\delta$. Furthermore, all simulation results emulate the random case, i.e. a job is sent to an open server selected uniformly at random.

\subsection{Baseline version of the \alg}

\begin{figure}[!tbp]
  \centering
  \begin{subfigure}[b]{\twopwidth\textwidth}
    \includegraphics[width=\columnwidth]{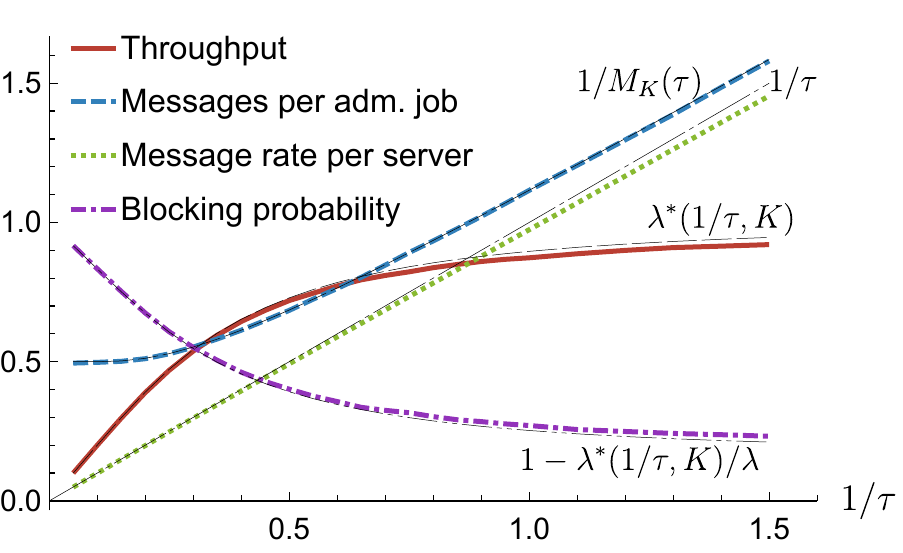}
    \caption{$K=2$.} \label{fig:d}
  \end{subfigure}
  \hfill
  \begin{subfigure}[b]{\twopwidth\textwidth}
    \includegraphics[width=\columnwidth]{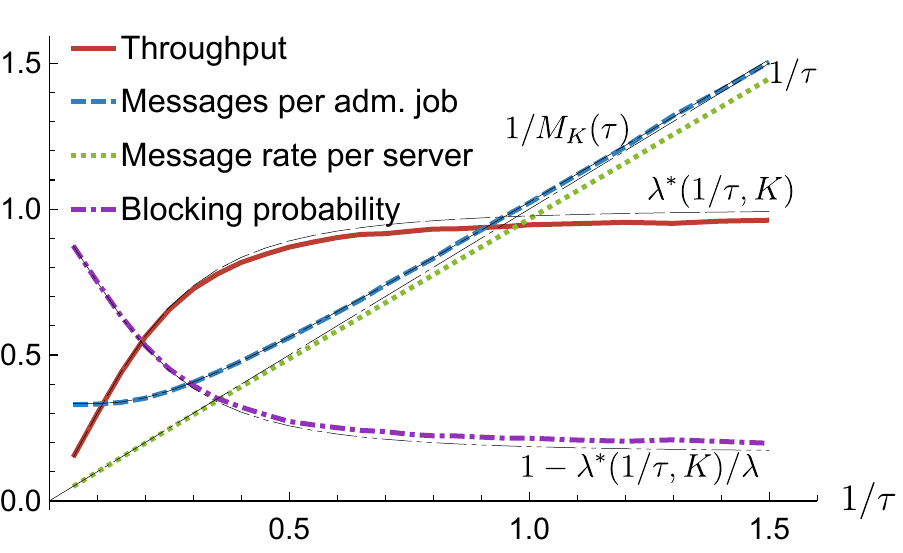}
    \caption{$K=3$.}\label{fig:d2}
  \end{subfigure}
  \caption{Simulation results for the \alg for $\lambda =1.2$ and $N=100$. Numerical values of the throughput bound $\lambda^*(1/\tau,K)$, the associated blocking probability bound $1-\lambda^*(1/\tau,K) / \lambda$, the average number of messages per admitted job $1/M_K(\tau)$ and $1/\tau$ are also shown with thin black lines.}
\end{figure}

First, we evaluate the \alg itself in Figures \ref{fig:d} and \ref{fig:d2} for $K=2$ and $K=3$ respectively. 
We note that the message rate stays below the line $y=1/\tau$, confirming that it never exceeds $1/\tau$. The throughput and blocking probability achieved by the \alg are nearly indistinguishable from the respective asymptotic values (upper and lower bounds, respectively), especially at lower and medium values of the communication budget $1/\tau$. For higher values of the communication budget, the throughput and blocking probability slightly diverge from the asymptotic values but remain remarkably close nevertheless. This demonstrates that the asymptotic optimality properties of the \alg as stated in Theorems \ref{th:gen} and \ref{th:asymopt} already manifest themselves in moderately large systems.

\newcommand{\aujsq}{$\textup{AUJSQ}^{det}(\delta)$\xspace}
In order to provide further insight in the asymptotic optimality, we compare the baseline version of the \alg with several variants and alternative scenarios that are not analytically tractable. \\

Specifically, in the next two subsections, we examine the following variants through simulations:
\begin{itemize}
\item ``non-idling''; open servers continue working, but convey their queue length as if they had not been working while being open,
\item ``work-conserving''; open servers continue working and convey their actual queue lengths at update epochs.
\end{itemize}

At first sight, one might suspect that these variants achieve a possibly larger throughput. As we will see however, the differences are small and are only observable at low load or in systems with few servers.

In Subsection \ref{subsec:aujsq} we make a comparison with the \aujsq scheme considered in \cite{BBL19}, which is not analytically tractable either but seems to be asymptotically throughput-optimal as well.

\subsection{Non-idling variant}

Open servers do not work on jobs in the baseline version of the \alg. While Theorem \ref{th:asymopt} showed that the forced idling does not affect the achieved throughput in large-scale systems, it is still interesting to investigate the consequences of this design. In the non-idling variant, open servers do work on jobs, but they convey their queue length to the dispatcher as if they had not been working on jobs while being labeled open. While this variant may seem fundamentally different, the information that the dispatcher has is exactly the same as in the baseline version: the sets of open servers and their respective states coincide in both scenarios, as long as jobs are sent to the same open server.

\begin{figure}[!tbp]
  \centering
  \begin{subfigure}[b]{\twopwidth\textwidth}
    \includegraphics[width=\columnwidth]{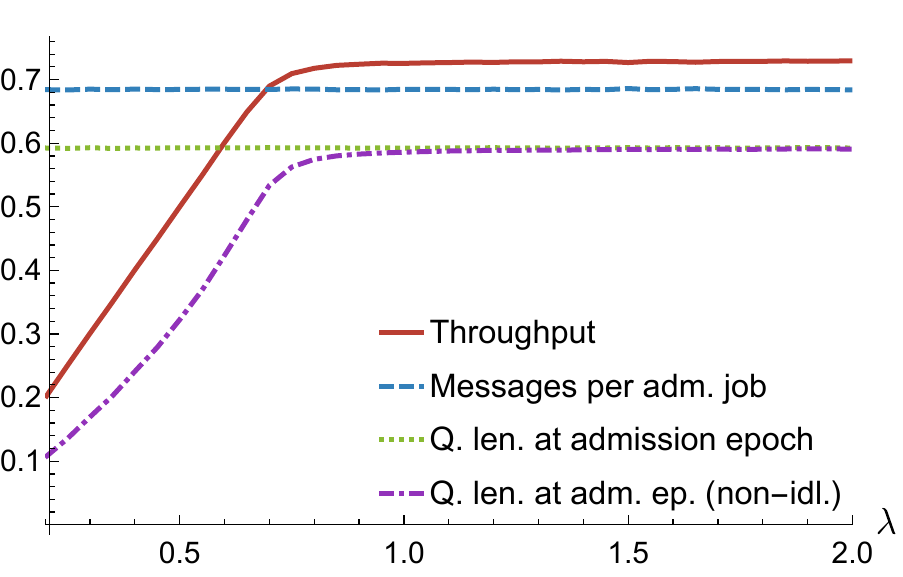}
    \caption{Non-idling variant.} \label{fig:L}
  \end{subfigure}
  \hfill
  \begin{subfigure}[b]{\twopwidth\textwidth}
    \includegraphics[width=\columnwidth]{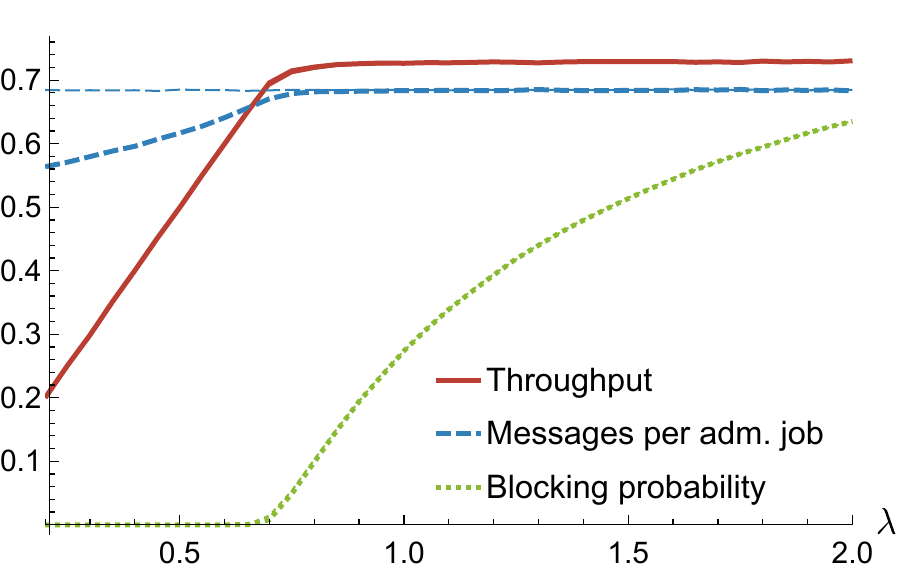}
    \caption{Work-conserving variant.}\label{fig:keepworking}
  \end{subfigure}
  \caption{Simulation results: comparison between two variants (thick lines) and the baseline scenario (thin lines), for $K=2$, $\tau=2$ and $N=500$, yielding a throughput bound $\lambda^*(1/2,2)\approx 0.73$.}
\end{figure}

In particular, the equilibrium distribution of the server states as provided in Proposition \ref{prop:productform} applies to the non-idling variant as well, and the throughput and the number of messages exchanged per admitted job are identical in both scenarios.
The only difference arises in the expected queue lengths encountered by admitted jobs: they are somewhat smaller in the non-idling scenario, as illustrated by the simulation results presented in Figure \ref{fig:L}.

At low load values, there are instants where there is time for servers to execute jobs when they are open. This causes a distinction between the two variants, since in the non-idling variant jobs join shorter queues. In Section \ref{sec:ext}, we will consider a tractable extension of the \alg that aims to reduce the queue lengths. As the number of servers grows however, an overflow of arrivals will cause open servers to have less time to execute jobs, which causes the queue lengths to be similar in both scenarios. This viewpoint provides further intuition why the \alg is still asymptotically optimal.

\subsection{Work-conserving variant}\label{subsec:wcd}
We now turn to a work-conserving variant of the \alg, in which open servers also work on jobs, and in fact convey their actual queue length at an update epoch. 
In this case the evolution of the server states is different, and the equilibrium distribution provided in Proposition \ref{prop:productform} no longer applies.

The throughput and blocking probability are similar in both scenarios. This may be intuitively explained as follows. When $\lambda\geq \lambda^*(1/2,2)$, Theorem \ref{th:asymopt} shows that there are hardly ever any open servers, and hence there should not be any substantial difference between the two variants, which is corroborated by Figure \ref{fig:keepworking}.

\begin{figure}[!tbp]
  \centering
  \begin{minipage}[b]{\twopwidth\textwidth}
    \includegraphics[width=\columnwidth]{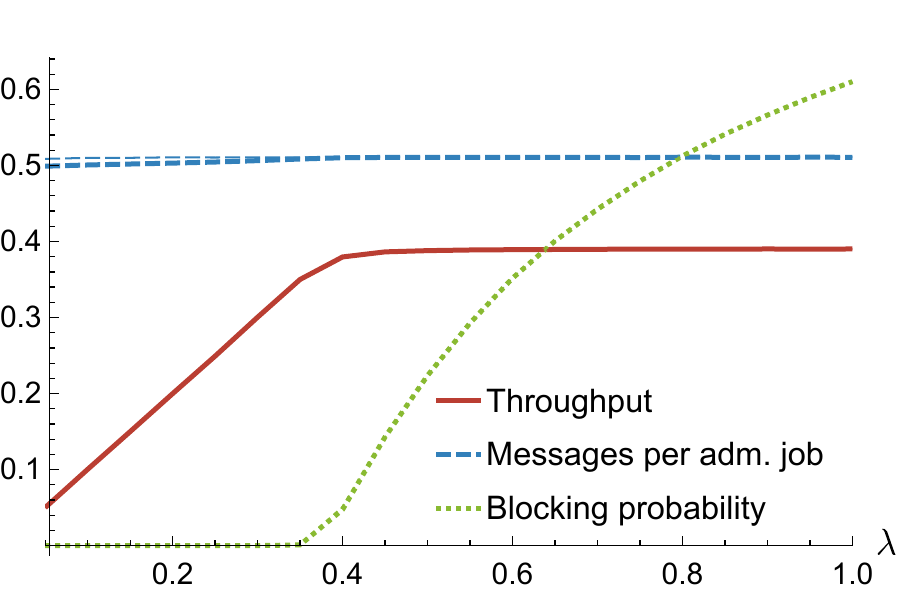}
    \caption{Simulation results: comparison between the baseline scenario (thin lines) and the work-conserving variant (thick lines) for $K=2$, $\tau=5$ and $N=500$, so that $\lambda^*(1/5,2)\approx 0.39$.} \label{fig:keepworking2}
  \end{minipage}
  \hfill
  \begin{minipage}[b]{\twopwidth\textwidth}
    \includegraphics[width=\columnwidth]{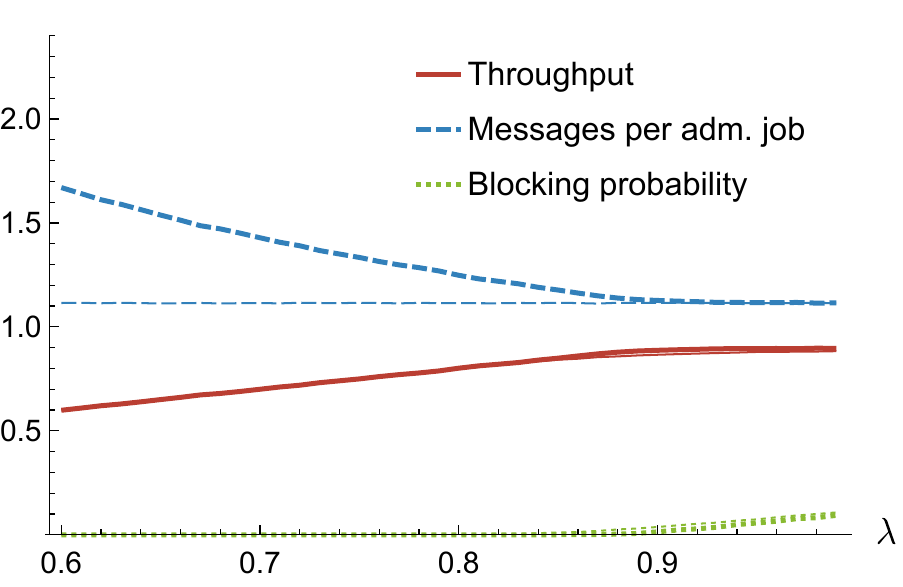}
    \caption{Simulation results: comparison between the baseline scenario (thin lines) and the \aujsq scheme (thick lines) for $K=2$, $\tau=1$ and $N=500$, so that $\lambda^*(1,2)\approx 0.90$.\\} \label{fig:fighjsq}
  \end{minipage}
\end{figure}

When $\lambda < \lambda^*(1/2,2)$, there can be a significant number of open servers. Theorem \ref{th:asymopt} however implies that the \alg approaches zero blocking and throughput $\lambda$ in this case. While it is plausible that the work-conserving variant achieves that as well, as attested by Figure \ref{fig:keepworking}, it is simply not feasible to achieve lower blocking or higher throughput. The only room for improvement is thus in the number of message exchanges per admitted job, and Figure \ref{fig:keepworking} demonstrates that the work-conserving variant indeed provides some gain compared to the \alg in that regard.
To put that observation in perspective, consider Corollary \ref{cor:comm}. As one can see, the communication overhead is strictly decreasing in $\tau$. For such a low arrival rate, the \alg permits to choose the update interval $\tau$ much larger. Figure \ref{fig:keepworking2} confirms that the choice $\tau=5$ largely eliminates the difference in communication overhead between the work-conserving variant and the baseline version.

\subsection{Comparison with the $\textup{AUJSQ}^{det}(\delta)$ scheme}\label{subsec:aujsq}
We now compare the \alg with the \aujsq scheme \cite{BBL19}, which is somewhat similar, except that every server is updated \emph{exactly} every $\tau = 1/\delta$ time units based on a timer. 
Thus the \aujsq scheme might update servers even when they are known to have strictly less than $K=2$ jobs in queue. 
There are further minor differences: in \aujsq jobs are assigned to the server with the lowest state (so giving preference to servers that are more likely to be empty) and open servers do work on jobs. In contrast to \cite{BBL19}, we consider a variant of the \aujsq scheme in which jobs are blocked when the dispatcher is not aware of any servers that are guaranteed to have strictly less than $K=2$ jobs in queue. The comparison is shown in Figure \ref{fig:fighjsq}.

It is important to note that in the \alg the expected number of messages \emph{per admitted job} is independent of $\lambda$, while in the \aujsq scheme the expected number of messages \emph{per time unit} is independent of $\lambda$. We observe that the average number of messages per admitted job coincides when $\lambda > \lambda^*(1/\tau,K)$. While it is natural to expect that the \aujsq scheme offers similar asymptotic optimality properties, it lacks the mathematical tractability of the \alg to facilitate a rigorous proof argument. 

\subsection{Non-exponential service times}
\begin{figure}[!tbp]
  \centering
  \begin{subfigure}[b]{\twopwidth\textwidth}
    \includegraphics[width=\columnwidth]{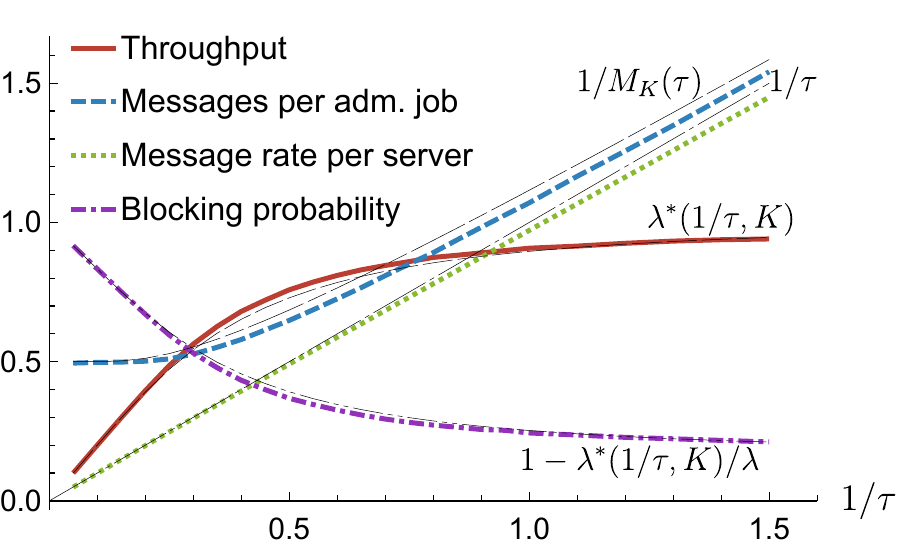}
    \caption{Gamma(2, 2).} \label{fig:dist1}
  \end{subfigure}
  \hfill
  \begin{subfigure}[b]{\twopwidth\textwidth}
    \includegraphics[width=\columnwidth]{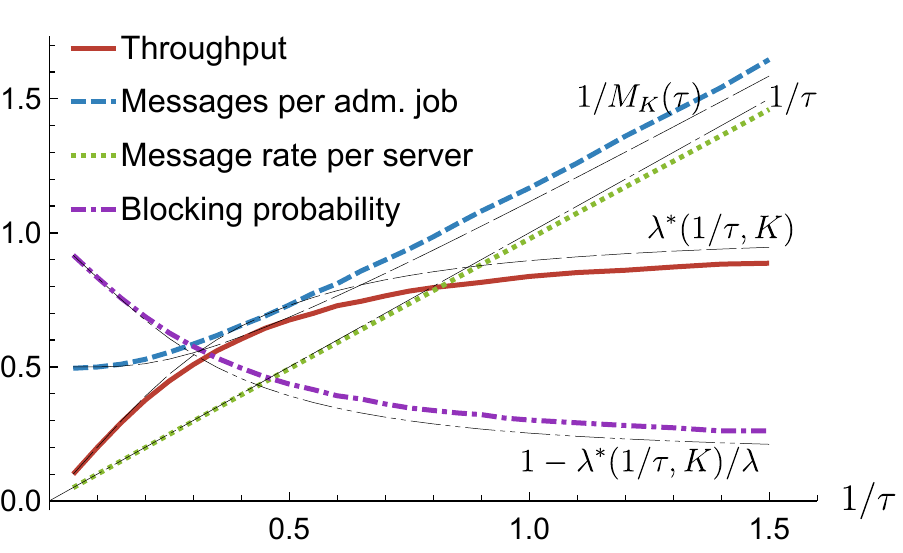}
    \caption{Gamma(1/2, 1/2).} \label{fig:dist2}
  \end{subfigure}
  \caption{Simulation results for the \alg for $K=2$, $\lambda =1.2$ and $N=100$, and non-exponential service times. Numerical values of the throughput bound $\lambda^*(1/\tau,K)$, the associated blocking probability bound $1-\lambda^*(1/\tau,K) / \lambda$, the average number of messages per admitted job $1/M_K(\tau)$ and $1/\tau$ are also shown with thin black lines.}
\end{figure}

We conclude our simulation experiments with analyzing the hyper-scalable scheme for non-exponential service time distributions. In Figure \ref{fig:dist1}, the service times are Gamma(2,2) distributed. The throughput of the hyper-scalable algorithm slightly exceeds $\lambda^*(1/\tau,K)$, the maximum throughput when job sizes are exponential. The number of messages per admitted job is also lower than $1/M_K(\tau)$. This is all explained by the fact that the tail of the Gamma(2,2) distribution is smaller than the tail of the exponential distribution. This means that more jobs are completed in a fixed time interval, which increases the effectiveness of the messages sent.
The service time distribution in Figure \ref{fig:dist2} is Gamma(1/2,1/2). The opposite effect is observed: the throughput is lower compared to Figure \ref{fig:d} and the message rate is larger, because of the heavier tail.

\section{Extension aimed at minimizing queue lengths} \label{sec:ext}\label{subsec:ext}
While the \alg is asymptotically throughput-optimal given the message rate $\delta$ and queue limit $K$, it does not make any explicit effort beyond that to minimize queue lengths or delays experienced by jobs.
Motivated by that observation, we now consider an extension of the \alg aimed at minimizing waiting times.
In this extension, a server that receives its $i$-th job after an update at which its queue length was $k$, becomes closed for $\tau_{k,k+i}$ time units. After this time, it becomes open if $k+i<K$ and it is updated if $k+i=K$. Thus, servers are not only closed when they become full, but are closed for a while after every job they receive. 

Henceforth, we focus on the case $K=2$ for the ease of exposition, and we set $\tau_{0,0}=0$, $\tau_{0,1}=\tau_{1,1}=\tau_1$, $\tau_{0,2}=\tau_{1,2}=\tau_2$ and $\tau_{2,2}=\tau_3$. We can put $\tau_{0,0}$ to zero without loss of generality as it makes no sense to have a cool-down period for an empty server. As a consequence there is no difference between servers that had zero jobs or one job at the previous update epoch, so we can set $\tau_{0,1} = \tau_{1,1}$, and $\tau_{0,2} = \tau_{1,2}$ as well. 
Let $p_{2j}$ be the probability that $j$ jobs remain after an update, when there were zero or one jobs just after the latest update epoch. This means that the server had $\tau_{0,1}$ time units to work on the first job and another $\tau_{0,2}$ time units after both jobs were dispatched to it. This gives
$
p_{20} = \ee^{-\tau_1} (1-\tau_2 \ee^{-\tau_2} - \ee^{-\tau_2}) + (1-\ee^{-\tau_1})(1-\ee^{-\tau_2})$, 
$p_{22} = \ee^{-\tau_1} \ee^{-\tau_2}$ and $p_{21}=1-p_{20}-p_{22}$. Let $q_{2j}$ be the probability that $j$ jobs remain after an update, when there were two jobs just after the latest update epoch. This gives
$q_{20} = 1-\ee^{-\tau_3} - \tau_3 \ee^{-\tau_3}$, 
$q_{22} = \ee^{-\tau_3}$ and $q_{21}=1-q_{20}-q_{22}$.

Servers can be in either of the five following states.
\begin{enumerate}
\item[$A_1$] The server is idle and open.
\item[$B_1$] The server had zero jobs during the previous update moment and received one job since, or the server had one job during the previous update moment and received no jobs since. The server is now marked closed for $\tau_1$ time units.
\item[$A_2$] The server had zero jobs during the previous update moment and received one job since, or the server had one job during the previous update moment and received no jobs since. The server was marked closed for $\tau_1$ but is now open.
\item[$B_2$] The server had zero jobs during the previous update moment and received two jobs since, or the server had one job during the previous update moment and received one job since. The server is now marked closed for $\tau_2$ time units.
\item[$B_3$] The server had two jobs during the previous update moment and is now marked closed for $\tau_3$ time units.
\end{enumerate}

\begin{figure}[t!]\centering
\includegraphics[width=0.7\columnwidth]{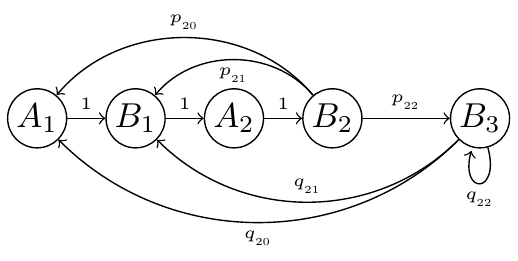}
\caption{Schematic representation of the server states and transitions when $K=2$.}\label{fig1}
\end{figure}

The transitions are schematically represented in Figure \ref{fig1}, with the transition probabilities as defined earlier.

The system dynamics under this extension of the hyper-scalable scheme can also be represented in terms of a closed queueing network with one single-server node that holds two classes of customers and three infinite-server nodes.
The states $A_1$ and $A_2$ correspond to the two classes that customers can be of when they are present at the single-server node. The states $B_1$, $B_2$ and $B_3$ each correspond to one of the three infinite-server nodes in the network, with deterministic service times $\tau_1$, $\tau_2$ and $\tau_3$, respectively.

\begin{myproposition} \label{prop:prodformext}
The equilibrium distribution of the system with $N$~servers is 
\begin{equation}\label{eq:eqdistext}
\begin{split}
&\pi(n_1,n_2,m_1,m_2,m_3) \\
&= H_N^{-1}
\frac{(n_1+n_2)!}{n_1! n_2!} 
\left(\frac{\pp_1}{\lambda N}\right)^{n_1} 
\left(\frac{\pp_2}{\lambda N}\right)^{n_2} 
\frac{(\qq_1 \tau_1)^{m_1}}{m_1!} 
\frac{(\qq_2 \tau_2)^{m_2}}{m_2!} 
\frac{(\qq_3 \tau_3)^{m_3}}{m_3!}
\end{split}
\end{equation}
if $n_1+n_2+m_1+m_2+m_3=N$, 
with $(\pp_1,\pp_2,\qq_1,\qq_2,\qq_3)=(p_{20}+\frac{p_{22} q_{20}}{1-q_{22}},1,$$1,$$1,$$\frac{p_{22}}{1-q_{22}})$ and normalization constant $H_N=$
\[
\sum_{v_1+v_2+w_1+w_2 + w_3 = N} \frac{(v_1+v_2)!}{v_1! v_2!} \left(\frac{\pp_1}{\lambda N}\right)^{v_1}
\left(\frac{\pp_2}{\lambda N}\right)^{v_2}
\frac{(\qq_1\tau_1)^{w_1}}{w_1!}
\frac{(\qq_2\tau_2)^{w_2}}{w_2!}
\frac{(\qq_3\tau_3)^{w_3}}{w_3!},
\]
where $n_i$ is the number of open servers in state $A_i$ and $m_i$ the number of closed servers in state $B_i$.
\end{myproposition}
The proof of Proposition \ref{prop:prodformext} is provided in Section \ref{sec:proof}.

The equilibrium distribution \eqref{eq:eqdistext} can be simplified when only the number of open and closed servers are counted, as shown in the next corollary.
\begin{mycorollary} \
\begin{itemize}
\item The equilibrium probability of there being $n$ open servers and $N-n$ closed servers under the extension equals
\[
\pi(n,N-n) = 
\frac{
\left(\frac{\gamma_1+\gamma_2}{\lambda N}\right)^n 
\frac{\left(\qq_1 \tau_1 +\qq_2 \tau_2 +\qq_3 \tau_3  \right)^{N-n}}{(N-n)!}
}{
\sum_{w=0}^N\left(\frac{\gamma_1+\gamma_2}{\lambda N}\right)^w \frac{\left( \qq_1 \tau_1 +\qq_2 \tau_2 +\qq_3 \tau_3  \right)^{N-w}}{(N-w)!}
}.
\]
In particular, because of the PASTA property, the blocking probability is given by
\[
\pi(0,N) = \frac{
\frac{ \left(x N \right)^N }{N!}
}
{
\sum_{w=0}^N \frac{\left(x N\right)^w}{w!}
},
\]
with $x = \lambda \times \frac{\qq_1 \tau_1 +\qq_2 \tau_2 +\qq_3 \tau_3}{\gamma_1+\gamma_2}$, and $\pi(0,N)\to \max\{0,1-\lambda^*(\tau_1,\tau_2,\tau_3)/\lambda\}
$
as $N \to \infty$, which equals zero when $\lambda \leq \lambda^*(\tau_1,\tau_2,\tau_3) :=  \frac{\pp_1+\pp_2}{\qq_1\tau_1+\qq_2\tau_2+\qq_3\tau_3}$.
\item
The average number of messages per admitted job equals
\[
u(\tau_1,\tau_2,\tau_3) := \frac{\qq_2+\qq_3}{\pp_1+\pp_2}.
\]
\item
The average queue position of an admitted job equals
\[
q(\tau_1,\tau_2,\tau_3) := \frac{\ee^{-\tau_1}}{\pp_1+\pp_2}.
\]
\end{itemize}
\end{mycorollary}

The last two statements follow directly from the relative throughput values.
These exact expressions for the maximum throughput $\lambda^*$, the average number of updates per admitted job $u$ and the average queue position $q$ of admitted jobs, allow us to evaluate the performance of this extension.

\begin{figure}[!tbp]
  \centering
  \begin{subfigure}[b]{\twopwidth\textwidth}
    \includegraphics[width=\columnwidth]{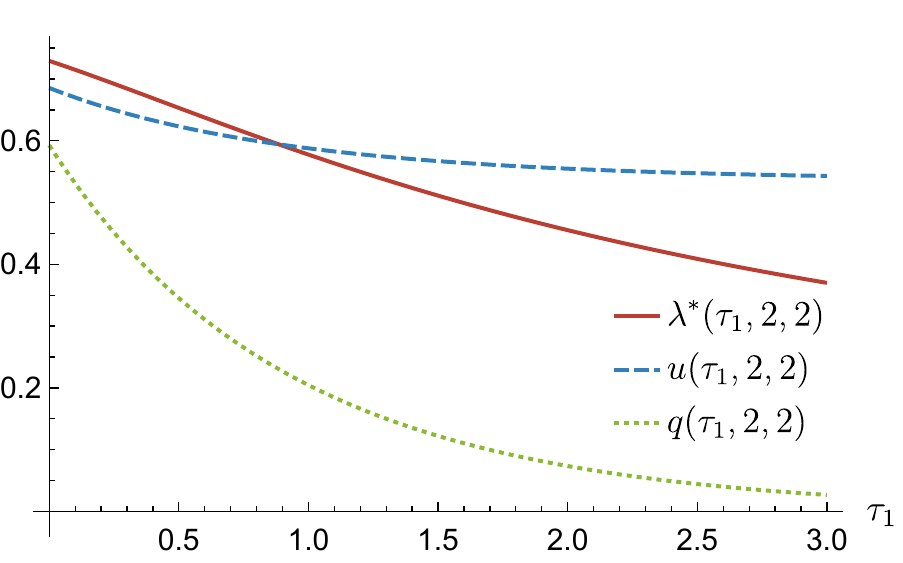}
    \caption{} \label{fig:ext1}
  \end{subfigure}
  \hfill
  \begin{subfigure}[b]{\twopwidth\textwidth}
    \includegraphics[width=\columnwidth]{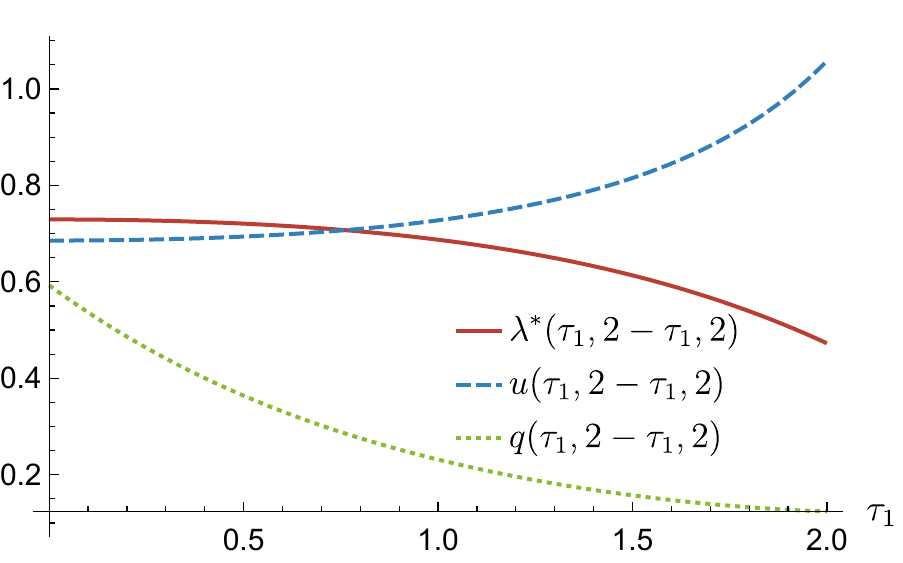}
    \caption{} \label{fig:ext4}
  \end{subfigure}
  \caption{Maximum throughput $\lambda^*$, average number of updates per admitted job $u$ and average queue position of admitted jobs $q$ as a function of $\tau_1$.}
\end{figure}

In Figure \ref{fig:ext1}, the value of $\tau_1$ is varied while the values of $\tau_2$ and $\tau_3$ are kept constant. Since $\tau_1$ represents the time that a server is closed when it has one job in queue, the result is that the second job that is sent to the server experiences a shorter queue in expectation. Indeed, for larger values of $\tau_1$, the mean experienced queue length $q$ decreases. As a further benefit, the mean number of updates decreases as well, since an idle server will take at least $\tau_1+\tau_2$ time units to be updated. The penalty incurred for these advantages is that the maximum throughput, $\lambda^*$, drops below the value of $\lambda^*(\delta,K)$ as asymptotically achieved by the baseline version of the \alg, since servers may become idle during the $\tau_1$ time in which they will not receive any more jobs.

Finally, in Figure \ref{fig:ext4} we show that a trade-off between the parameters is possible. $\tau_1$ is increased while $\tau_2$ is decreased, and this leads to interesting behavior. Around the point $\tau_1=0$, the values of $\lambda^*$ and $u$ do not change when the parameters are altered, but the value of $q$ does change. Such a trade-off might be worth it in scenarios where mean queue lengths play an important role.

\section{Closed queueing network and further proofs}\label{sec:proof}\label{sec:cqn}
In this section we establish product-form distributions for a general closed queueing network scenario which captures the network representations of the \alg and the extension considered in the previous section as special cases. This provides the proofs of Propositions \ref{prop:productform} and \ref{prop:prodformext}.

The closed queueing network consists of $N$ \customers circulating among one single-server node and $B$ infinite-server nodes. \Customers can be of $A$ classes while at the single-server node, denoted by $A_1,\hdots, A_A$. Denote the infinite-server nodes by $B_1,\hdots, B_B$. The routing probabilities are denoted by $\PP_{x \to y}$; this is the probability that a \customer transitions from $x$ to $y$ ($x$ and $y$ may correspond to either a class or an infinite-server node). 

Service completions at the multi-class single-server node occur at an exponential rate $\lambda N$. The \customer that completes service is either selected uniformly at random, or in a FCFS manner, where the next \customer is the one that transitioned last. If the selected \customer is of class $i$, then it immediately returns to the single-server node as a class-$j$ \customer with probability $\PP_{A_i \to A_j}$ or it moves to node $B_j$ with probability $\PP_{A_i\to B_j}$. The service times at the infinite-server node $B_i$ are deterministic and equal to $\tau_i$. Upon completing service at node $B_i$, a customer moves to the single-server node as a class-$j$ customer with probability $\PP_{B_i\to A_j}$, or to node $B_j$ with probability $\PP_{B_i \to B_j}$.

The relative throughput values may be calculated from the traffic equations,
\[
\begin{cases}
\pp_i = \sum_{j=1}^A \PP_{A_j \to A_i} \times \pp_j + \sum_{j=1}^B \PP_{B_j \to A_i} \times \qq_j, \\
\qq_i = \sum_{j=1}^A \PP_{A_j \to B_i} \times \pp_j + \sum_{j=1}^B \PP_{B_j \to B_i} \times \qq_j,
\end{cases}
\]
where $\pp_i$ stands for the relative throughput of class $A_i$ at the single-server node and $\qq_i$ for the relative throughput at node $B_i$. We assume a ``single-chain network'', where the routing probability matrix is irreducible, meaning that all customers can reach all classes and nodes.

\begin{myproposition}\label{prop:cqn} \
\begin{enumerate}[label=(\alph*)]
\item \label{th:cqn1}
The equilibrium distribution of the system with $N$~\customers is 
\begin{equation}\label{eqth1}
\begin{split}
&\pi(n_1,n_2,\hdots,n_A,m_1,m_2,\hdots,m_B) \\&= 
F_N^{-1}
\frac{(n_1+\hdots+n_A)!}{n_1! \cdots n_A!} 
\prod_{i=1}^A
\left(\frac{\pp_i}{\lambda N}\right)^{n_i}
\prod_{j=1}^B 
\frac{(\qq_j \tau_j)^{m_j}}{m_j!}
\end{split}
\end{equation}
if $n_1+\hdots+n_A+m_1+\hdots+m_B=N$, with normalization constant
\[
F_N = 
\sum_{v_1+\hdots+v_A+w_1+\hdots+w_B = N} \frac{(v_1+\hdots+v_A)!}{v_1! \cdots v_A!} 
\prod_{i=1}^A
\left(\frac{\pp_i}{\lambda N}\right)^{v_i}
\prod_{j=1}^B 
\frac{(\qq_j \tau_j)^{w_j}}{w_j!},
\]
where $n_i$ is the number of \customers of class $A_i$ at the single-server node and $m_j$ the number of \customers at infinite-server node $B_j$.

\item \label{th:cqn2} 
The equilibrium probability of there being $n$ customers at the single-server node and $N-n$ customers in total at all the infinite-server nodes equals
\begin{equation}\label{eq:th3gen}
\begin{split}
\pi(n, N-n) &= \sum_{
\substack{n_1+\hdots+n_A=n\\ m_1+\hdots+m_B=N-n }} \pi(n_1,\hdots,n_A,m_1,\hdots,m_B  ) \\&=  
\frac{
\left(\frac{\sum_{i=1}^A \pp_i}{\lambda N}\right)^n 
\frac{\left(\sum_{j=1}^B \qq_j \tau_j \right)^{N-n}}{(N-n)!}
}{
\sum_{w=0}^N\left(\frac{\sum_{i=1}^A \pp_i}{\lambda N}\right)^w \frac{\left(\sum_{j=1}^B \qq_j \tau_j \right)^{N-w}}{(N-w)!}
}.
\end{split}
\end{equation}

In particular, with $R=\frac{\pp_1+\hdots+\pp_A}{\qq_1 \tau_1+\hdots + \qq_B \tau_B}$ and $x=\lambda/R$, because of the PASTA property, the probability that no customer resides at the single-server node is
\[\pi(0,N) = \frac{
\frac{ \left(x N \right)^N }{N!}
}
{
\sum_{w=0}^N \frac{\left(x N\right)^w}{w!}
},
\]
and
$
\pi(0,N) \to \max\{0, 1-R/\lambda\}
$
as $N \to \infty$ which equals zero when $\lambda \leq R$.
\end{enumerate}
\end{myproposition}

In order to prove Proposition \ref{prop:cqn}, we will verify that the equilibrium distribution \eqref{eqth1} satisfies the balance equations of the closed queueing network.

\subsection{Proof of Proposition \ref{prop:cqn}}
In order to verify the balance equations, we may assume that the service times of the infinite-server nodes are exponentially distributed even though in our closed queueing network, the service times are deterministic. This is because the equilibrium distribution \eqref{eqth1} is insensitive to the service time distribution of nodes and only depends on the means of them (see Chapter 3 of \cite{K11} for a further discussion on this). 

To see this, consider one infinite-server node $D$ with exponential service rate $\mu_D$ and throughput value $\qq_D$. This node adds the term 
\begin{equation}\label{eq:ins}
\frac{(\qq_D / \mu_D)^d }{ d!}
\end{equation}
to the product-form equilibrium distribution, representing the presence of $d$ customers in the infinite-server node. We now replace this infinite-server node by a series of infinite-server nodes, denoted by $D_1, \hdots, D_M$, each with an exponential service rate $M \mu_D$. The transition probabilities are altered in such a way that every transition previously to node $D$, now transitions to node $D_1$ instead. Customers then transition from node $D_i$ to $D_{i+1}$ for $i=1,\hdots,D-1$ with probability one. Finally, any transition previously from node $D$, will now transition from node $D_M$. This construction makes every customer stay in this collection of nodes for $M$ exponentially distributed phases, which is an $\textup{Erlang}(M,M \mu)$ distributed random variable. All other throughput values in the network remain equal.

The throughput values of all these nodes will be equal to $\qq_D$ (since they are in series). Finally, similarly to the simplification of \eqref{eqth1} to \eqref{eq:th3gen}, all nodes $D_1, \hdots, D_M$ may be aggregated, which would lead to a term 
\[
\frac{(\sum_{i=1}^M \qq_D / (M\mu_D))^d}{d!} = \frac{(\qq_D / \mu_D)^d }{ d!}
\]
in the equilibrium probability, representing the presence of $d$ customers in total in the infinite-server nodes $D_1,\hdots, D_M$. Note that the term in the RHS exactly matches the term \eqref{eq:ins}, that appears when the node $D$ has an exponentially distributed service time.
This shows that the equilibrium distribution does not change when an exponential node is replaced by an $\textup{Erlang}(M,M\mu)$ node, for any integer $M$, which can also be verified by substitution in the balance equations. Of course, each infinite-server node $B_i$ with $\mu_i=1/\tau_i$ can be replaced by such an Erlang distribution using this construction. 

Because an $\textup{Erlang}(M,M\mu)$ random variable converges to a deterministic quantity $1/\mu$ as $M$ tends to infinity, this indicates that the equilibrium distribution also holds with infinite-server nodes that have deterministic service times. In fact, the node $D$ may be replaced by any phase-type distribution, and every distribution may be approximated arbitrarily closely by phase-type distributions, implying that the equilibrium distribution in \eqref{eqth1} in fact holds for generally distributed service times with mean $\tau_i$ at the infinite-server node $B_i$ as well, although that is not directly relevant for our purposes. \\

We will now verify that \eqref{eqth1} indeed solves the balance equations for the random order of service case, and we will use $\mu_i = 1/ \tau_i$, representing the rates of the infinite-server nodes. The proof for the FCFS case is quite similar, but involves a more detailed state representation, and is deferred to Appendix \ref{app:proof}.
\begin{proof}[Proof of part \ref{th:cqn1} - ROS]
For conciseness, denote by $(a,b)$ the vector $(a_1,$$\hdots,a_A,$ $b_1,\hdots,b_B)$ and by $e_i$ the $i$-th unit vector.

Note that \eqref{eqth1} is a proper distribution by definition. Since the equilibrium distribution is unique, it suffices to verify that \eqref{eqth1} satisfies the following set of balance equations:
\[
\begin{split}
&\left(\ind{a_1+\hdots+a_A>0} \lambda N+ b_1 \mu_1 +\hdots+ b_B \mu_B \right) \pi(a,b) \\
&= 
\sum_{i=1}^A \sum_{j=1}^A \ind{a_j > 0} \PP_{A_i \to A_j} 
\frac{a_i + \ind{i\neq j}}{a_1+\hdots+a_A} \lambda N \pi(a+e_i-e_j,b)\\
&+\sum_{i=1}^A \sum_{j=1}^B \ind{b_j > 0} \PP_{A_i \to B_j}
\frac{a_i + 1}{a_1+\hdots+a_A + 1} \lambda N \pi(a+e_i,b-e_j)\\
&+\sum_{i=1}^B \sum_{j=1}^A \ind{a_j > 0} \PP_{B_i \to A_j}
({b_i + 1})\mu_i \pi(a-e_j,b+e_i)\\
&+\sum_{i=1}^B \sum_{j=1}^B \ind{b_j > 0} \PP_{B_i \to B_j}
({b_i+\ind{i\neq j}})\mu_i \pi(a,b+e_i-e_j).
\end{split}
\]
The first line of the RHS refers to transitions where a \customer at the single-server node transitions to the same node and may change class. The second line refers to transitions from the single-server node to one of the infinite-server nodes. Lines three and four correspond to transitions from a infinite-server node, to the single-server node or to another infinite-server node, respectively. 

We will show that \eqref{eqth1} satisfies the balance equations. 
By using the definition of \eqref{eqth1} in the RHS, we obtain
\[
\begin{split}
&\sum_{i=1}^A \sum_{j=1}^A \ind{a_j > 0} \PP_{A_i \to A_j}
 \frac{a_i + 1}{a_1+\hdots+a_A} \lambda N \pi(a,b) \frac{a_j}{a_i+1}\frac{\pp_i}{\lambda N}\frac{\lambda N}{\pp_j}\\
+&\sum_{i=1}^A \sum_{j=1}^B \ind{b_j > 0} \PP_{A_i \to B_j}
 \frac{a_i + 1}{a_1+\hdots+a_A + 1} \lambda N \pi(a,b) \frac{a_1+\hdots+a_A+1}{a_i+1}\frac{\pp_i}{\lambda N}\frac{b_j}{\qq_j T_j}\\
+&\sum_{i=1}^B \sum_{j=1}^A \ind{a_j > 0} \PP_{B_i \to A_j}
 ({b_i + 1})\mu_i \pi(a,b) \frac{a_j}{a_1+\hdots+a_A}\frac{\lambda N}{\pp_j}\frac{\qq_i  \tau_i}{b_i+1}\\
+&\sum_{i=1}^B \sum_{j=1}^B \ind{b_j > 0} \PP_{B_i \to B_j}
 ({b_i+1})\mu_i \pi(a,b) \frac{\qq_i / \mu_i}{b_i+1}\frac{b_j}{\qq_j / \mu_j}.\\
\end{split}
\]
Next, we combine the inside sums, resulting in
\[
\begin{split}
&\sum_{j=1}^A \ind{a_j > 0} \frac{a_j}{a_1+\hdots+a_A} \frac{\lambda N}{\pp_j} \left[ \sum_{i=1}^A \PP_{A_1\to A_j} \pp_i + \sum_{i=1}^B \PP_{B_i\to A_j} \qq_i\right] \pi(a,b)\\
+&\sum_{j=1}^B \ind{b_j > 0} \frac{b_j}{\qq_j / \mu_j} \left[ \sum_{i=1}^A \PP_{A_i \to B_j} \pp_i + \sum_{i=1}^B \PP_{B_i \to B_j} \qq_i \right] \pi(a,b)\\
&=\left[ \sum_{j=1}^A \ind{a_j > 0} \frac{a_j}{a_1+\hdots+a_A} \lambda N + \sum_{j=1}^B {b_j}\mu_j \right] \pi(a,b) \\
&= \left(\ind{a_1+\hdots+a_A>0} \lambda N+ \sum_{j=1}^B {b_j}\mu_j \right) \pi(a,b).
\end{split}
\]
\end{proof}

\section{Conclusion}\label{sec:concl}
We established a universal upper bound for the achievable throughput of any dispatcher-driven algorithm for a given communication budget and queue limit.
We also introduced a specific hyper-scalable scheme which can operate at any given message rate and enforce any given queue limit, while allowing the system dynamics to be captured via a closed product-form network.
We leveraged the product-form distribution to show that the bound is tight, and that the proposed hyper-scalable scheme provides asymptotic optimality in the three-way trade-off among performance, communication and throughput.
Extensive simulation experiments were presented to illustrate the results and make comparisons with various alternative design options.

The work-conserving variant covered in Subsection~\ref{subsec:wcd} is especially worth discussing further.
Intuitively, letting servers work all the time seems better than pausing the servers when they become open, but this remains to be rigorously proven.


The extension aimed at minimizing waiting times that was introduced in
Section~\ref{sec:ext} warrants further attention as well.
For the baseline scenario, we were able to prove a strict relationship between the amount of communication and the throughput.
Likewise, there might exist a result, similar in spirit to Theorem~\ref{th:gen}, which provides an upper bound for the throughput \emph{and} the average queue position of admitted jobs, given a certain communication budget.
The main point of concern in this regard is that the concavity argument no longer seems to hold.

Finally, it would be worth investigating whether the current framework could be broadened further.
It may be possible for example to extend the category of algorithms considered, specifically allowing for pull-based schemes.
While the results in~\cite{BBZ20} imply that Theorem~\ref{th:gen} does not hold for pull-based schemes, there might be a larger upper bound covering such algorithms as well.
For further extensions, other performance metrics might be considered too, such as the mean waiting time as opposed to the throughput subject to a queue limit.

\section*{Acknowledgments}
This work is supported by the NWO Gravitation Networks grant 024.002.003 and an ERC Starting Grant. We would like to thank C\'{e}line Comte and Martin Zubeldia for several helpful discussions and suggestions.

\bibliography{biblioINI}{}
\bibliographystyle{plain}

\appendix

\section{Proof of Proposition \ref{prop:cqn} - FCFS case} \label{app:proof}

\begin{proof}[Proof of part \ref{th:cqn1} - FCFS]
The proof for the FCFS case consists of multiple steps. First we define a more detailed state space.
A state is represented by $(c,b)=((c_1,\hdots,c_m),(b_1,\hdots,b_B))$, which represents the situation where $m$ \customers are at the single-server node, and the order of the classes of \customers is saved as well: the $k$th \customer at the single-server node has class $c_k$. We will sometimes refer to the number of \customers of a specific class with $a_i = \sum_j \ind{c_j=A_i}$. Furthermore, $b_i$ customers are at the infinite-server node $B_i$. 

\emph{Equilibrium distribution for the extended state space.}
We will show that the equilibrium distribution (modulo normalization constant) of state $(c,b)$ equals
\begin{equation}\label{eq:eqproborder}
\tilde \pi(c,b) = \left(\frac{\pp_1}{\lambda N}\right)^{a_1} \cdots
\left(\frac{\pp_A}{\lambda N}\right)^{a_A} 
\frac{(\qq_1 / \mu_1)^{b_1}}{b_1!} \cdots
\frac{(\qq_B / \mu_B)^{b_B}}{b_B!}
\end{equation}
with $a_k$ the number of \customers of class $A_k$.

We assume FCFS arrivals of customers at the single-server node: \customers arrive at the end of the line at the single-server node and only the \customer first in line is able to transition.

\emph{Balance equations.}
First, we introduce the balance equations, in which the symbol $m$ is used to denote the length of vector $c$,
\begin{equation}\label{eq:balance}
\begin{split}
&\left(\ind{m>0} \lambda N+ {b_1}\mu_1+\hdots+ {b_B}{\mu_B}\right)  \tilde\pi(c,b) \\
&=
\sum_{i=1}^A \ind{m>0} \PP_{A_i\to c_m} \lambda N \tilde\pi((A_i,c_1,\hdots,c_{m-1}),b) \\
&+ \sum_{i=1}^A \sum_{j=1}^B \ind{b_j > 0} \PP_{A_i\to B_j} \lambda N \tilde\pi((A_i,c_1,\hdots,c_m), b - e_j ) \\
&+ \sum_{i=1}^B \ind{m>0} \PP_{B_i \to c_m} ({b_i+1})\mu_i  \tilde\pi((c_1,\hdots,c_{m-1}),b+e_i) \\
&+ \sum_{i=1}^B \sum_{j=1}^B \ind{b_j>0} \PP_{B_i \to B_j} ({b_i+\ind{i\neq j}})\mu_i \tilde\pi(c,b+e_i-e_j).
\end{split}
\end{equation}

The term before $\pi(c,b)$ on the LHS represents the outgoing rate of state $(c,b)$, which equals a rate of $\lambda N$ for the single-server node (if at least one \customer is present there) plus a rate of $b_j\mu_j$, for each infinite-server node $B_j$.

On the RHS, four possible transitions to state $(c,b)$ are shown preceded by the rate of the transitions. 
First, a transition from the non-empty single-server node makes the then first \customer change its class from $c_{m-1}$ to class $c_m$. If the previous class order at the single-server node is $c_m-1,c_1,\hdots, c_{m-1}$, then a transition to that node will make the class order exactly $c$. 
Second, if the previous class order at the single-server node is $K-1, c_1, \hdots, c_m$, then a transition from that node to a infinite-server node will make the class order exactly $c$. Additionally, if the number of \customers at infinite-server node $B_j$ was $b_j-1$, then it will become $b_j$ as the infinite-server node receives an extra \customer.
Third, any of the customers at the infinite-server nodes might transition to the single-server node.
Finally, customers might transition from and to one of the infinite-server nodes.

We will show that $\tilde \pi$ satisfies the balance equations.
By using the definition of $\tilde \pi$ in the RHS, we obtain
\begin{equation}\label{eq:balancefcfs}
\begin{split}
&\sum_{i=1}^A \ind{m>0} \PP_{A_i\to c_m} \lambda N \tilde\pi(c,b)
\frac{\pp_i}{\lambda N}\frac{\lambda N}{\pp_{c_m}} \\
+& \sum_{i=1}^A \sum_{j=1}^B \ind{b_j > 0} \PP_{A_i\to B_j} \lambda N \tilde\pi(c,b)
\frac{\pp_i}{\lambda N} \frac{b_j}{\qq_j \tau_j}\\
+& \sum_{i=1}^B \ind{m>0} \PP_{B_i \to c_m} ({b_i+1})\mu_i \tilde\pi(c,b) \frac{\lambda N}{\pp_{c_m}}
\frac{\qq_i / \mu_i}{b_i+1}\\
+& \sum_{i=1}^B \sum_{j=1}^B \ind{b_j>0} \PP_{B_i \to B_j} ({b_i+1})\mu_i \tilde\pi(c,b)
\frac{\qq_i / \mu_i}{b_i +1} \frac{b_j}{\qq_j / \mu_j}.
\end{split}
\end{equation}
Next, we reorganize terms, yielding
\begin{equation}
\begin{split}
&\ind{m>0}  \frac{\lambda N}{\pp_{c_m}} \tilde\pi(c,b) \left[ \sum_{i=1}^A \PP_{A_i\to c_m} \pp_i + \sum_{i=1}^B \PP_{B_i \to c_m} \qq_i \right]\\
&+\sum_{j=1}^B \ind{b_j>0} \frac{b_j}{\qq_j / \mu_j} \tilde\pi(c,b) \left[\sum_{i=1}^A \PP_{A_i \to B_j} \pp_i + \sum_{i=1}^B \PP_{B_i \to B_j} \qq_i \right]\\
&= \left[\ind{m>0} \lambda N   + \sum_{j=1}^B {b_j}\mu_j \right]\tilde\pi(c,b),
\end{split}
\end{equation}
which shows that $\tilde \pi$ is the equilibrium distribution of the extended state space.

Finally, note that in the original state space, only the number of customers of certain classes is tracked. Thus, $\pi(a,b)$ is an enumeration of $\tilde \pi(c,b)$ over all possible orders with the correct number of customers of a certain class. The number of possible orders is 
$\binom{a_1+\hdots+a_A}{a_1 \dots a_A}$,
which leads to $\pi(a,b) = \binom{a_1+\hdots+a_A}{a_1 \dots a_A} \pi(c,b)$; the description of $\pi$ as presented in the statement of the proposition.
\end{proof}

\end{document}